\documentclass[jcp,amsmath,amssymb,showkeys,floatfix,
reprint
]{revtex4-2}

\usepackage{bibentry}
\usepackage{appendix}
\usepackage[framemethod=tikz]{mdframed}
\mdfsetup{
	outerlinewidth=1pt,
	linecolor=white!90!black,
	innertopmargin=6pt,
	innerbottommargin=6pt,
	leftmargin=2pt,
	rightmargin=2pt
}
\usepackage[utf8]{inputenc}
\usepackage{float}

\usepackage{color}

\usepackage{graphicx}
\usepackage{dcolumn}
\usepackage{bm}
\usepackage{xfrac}
\usepackage{colortbl}

\usepackage{natbib}
\usepackage{bibentry}
\usepackage{float}
\usepackage{sidecap}
\usepackage[section]{placeins}
\usepackage{hyperref}
\hypersetup{
	linkcolor=blue,
	citecolor=blue,
	filecolor=blue,
	urlcolor=blue,
	colorlinks=true
}

\usepackage[caption=false]{subfig}

\usepackage{amsmath}
\usepackage{amssymb}
\usepackage{physics}
\usepackage{marvosym}
\usepackage{wasysym}
\usepackage{pifont}
\usepackage{wrapfig}
\usepackage{xifthen}
\usepackage{enumerate}
\usepackage{cancel}
\usepackage{MnSymbol}
\usepackage{mathtools}
\usepackage{relsize}

\usepackage{threeparttable}

\usepackage[normalem]{ulem}

\begin{document}

\title{ Large-scale sparse wavefunction circuit simulator for applications with the variational quantum eigensolver}
\author{J.~Wayne~Mullinax }
\affiliation{KBR, Inc., Intelligent Systems Division, Mail Stop 269-3, NASA Ames Research Center, Moffet Field, California  94035, USA}
\author{ Norm~M.~Tubman}
\email{norman.m.tubman@nasa.gov}
\affiliation{NASA Ames Research Center, Moffett Field, CA 94035 , USA}
\date{\today}
\begin{abstract}

The standard paradigm for state preparation on quantum computers for the simulation of physical systems in the near term has been widely explored with different algorithmic methods.
One such approach is the optimization of parameterized circuits, but this becomes increasingly challenging with circuit size.
As a consequence, the utility of large-scale circuit optimization is relatively unknown.
In this work we demonstrate that purely classical resources can be used to optimize quantum circuits in an approximate but robust manner such that we can bridge the resources that we have from high performance computing and see a direct transition to quantum advantage.  We show this through sparse wavefunction circuit solvers, which we detail here, and 
demonstrate a region of efficient classic simulation.
With such tools, we can avoid the many problems that plague circuit optimization for circuits with hundreds of qubits using only practical and reasonable classical computing resources.  
These tools allow us to 
probe the true benefit of variational optimization approaches on quantum computers, thus opening the window to what can be expected with near term hardware for physical systems.  We demonstrate this with a unitary coupled cluster ansatz on various molecules up to 64 qubits with tens of thousands of variational parameters.  

\end{abstract}

\maketitle

\section{Introduction}
\label{sec:introduction}
 One of the most promising applications of quantum computing is simulating physical systems \cite{Smith:2019:106,Bauer:2020:12685,McArdle:2020:015003}.   Many variational algorithms have been proposed, including the variational quantum eigensolver (VQE), which have been shown to have a wide range of applicability~ \cite{Peruzzo:2014:4213,McClean:2016:023023,Stavenger2022,Bassman2022,Magann2023,Luo2022,Huembeli2021,Aram2021,Arrasmith2021,Anastasiou2022,Carrasquilla2021,Martyn2022,Burton2022,Claudino2020,Tang2021,Chamaki2022,Romero2018,Ayral2022,Tilly2022,Cerezo2021,Luo2022,Guerreschi2020,McClean2017,Smelyanskiy2016,Cao2019,Smith2022,Lu2022,Zhao2020,Pathak2022,Kandala2017,Chamaki2022-1,Lin2020,Unmuth2022,Hug2020,Sherbert2021,Sherbert2022,Traps2022,Oganov2020,Miro2020,Kirby2021,Kirby2019,Li2021,Jahin2022}.   Variational approaches have many unknowns in terms of expected performance beyond small problems~\cite{Cao2019,Tilly2022,Cerezo2021}.   
 As system sizes are increased and we attempt to find quantum advantage over classical simulations, it remains an open question as to how variational algorithms will perform.  
Recent simulations have been able to test the algorithmic efficiency of a variety of different quantum algorithms~\cite{Hogg:2000:181,Farhi:2014,Harrigan:2021:332,AspuruGuzik:2005:1705,Hauke:2020:054401,Kitaev:1997:1191,Abrams:1997:2586,Abrams:1999:5162,Ang2022} on fairly large systems using novel approaches converted from classical simulations which include adiabatic state preparation \cite{Kremenetski2021-1}, QAOA \cite{Kremenetski:2021}, phase estimation \cite{Tubman:2020:2139} and variational quantum phase estimation \cite{Klymko:2022:020323}, among others. 
These algorithms use an approximation of a  sparse wavefunction using similar techniques that have been recently developed and expanded upon in the quantum chemistry community~\cite{Tubman2018}. 
Here, we take a recent algorithm developed for classical simulations of unitary coupled cluster \cite{Chen:2021:841}, and we  generalize it to sparse circuit simulations to demonstrate that it can be used to determine parameters for state preparation.  With this we can start exploring several exciting aspects of chemistry, physics, and material science applications on quantum hardware, which include:  1) Realistic benchmarking and testing of different circuit ansatze for physical problems beyond 20 qubits~\cite{Grimsley:2020:1} 2) A new paradigm in VQE algorithms for which much of the work in optimizing variational parameters is shifted to classical hardware (possible with high performance computing resources), thus reducing the amount of optimization on the expensive quantum hardware  3) Using approximate circuit simulators to reduce issues related to barren plateaus and mitigate other difficult aspects of optimizing on noisy hardware, thus expanding on what is possible with near term quantum hardware.   

The problems encountered in optimization on quantum hardware share some similarity with classical optimization problems, but some problems are unique.  While large scale optimization has many applications on classical hardware, the idea of barren plateaus~\cite{McClean:2018:4812,Traps2022,Arrasmith2021} has  largely been identified as a problem by communities looking at optimization on quantum hardware. Here we consider approaches in which high precision classical simulations can be used to find regions of convergence, and afterwards quantum hardware can be used to refine the optimization. We see this new paradigm (which has also been considered in other context~\cite{Ayral2022}) as a  path to demonstrate quantum advantage in the future for molecular simulations employing hundreds of qubits.    The sparse solver is key to this approach, as current exact circuit simulators can handle approximately 30 qubits ~\cite{Cao:2022:062452,McClean2017,Zhao2022,Smelyanskiy2016,Qiskit}.  The bottleneck for large scale optimizations is due to the enormous number of variational parameters and the size of the wavefunction when exact simulators are used. The simulations presented in this manuscript were run on a single computational node and do not need to utilize large scale parallelization. Although our approach is approximate, we are able to investigate the convergence behavior of the electronic energy while using double-zeta quality basis sets for up to 64 qubits, a  starting point for understanding the performance of wavefunction optimization with non-trivial basis sets. 


\section{Results and Discussion}
\label{sec:results}
\begin{figure}
\includegraphics[width=3.4in]{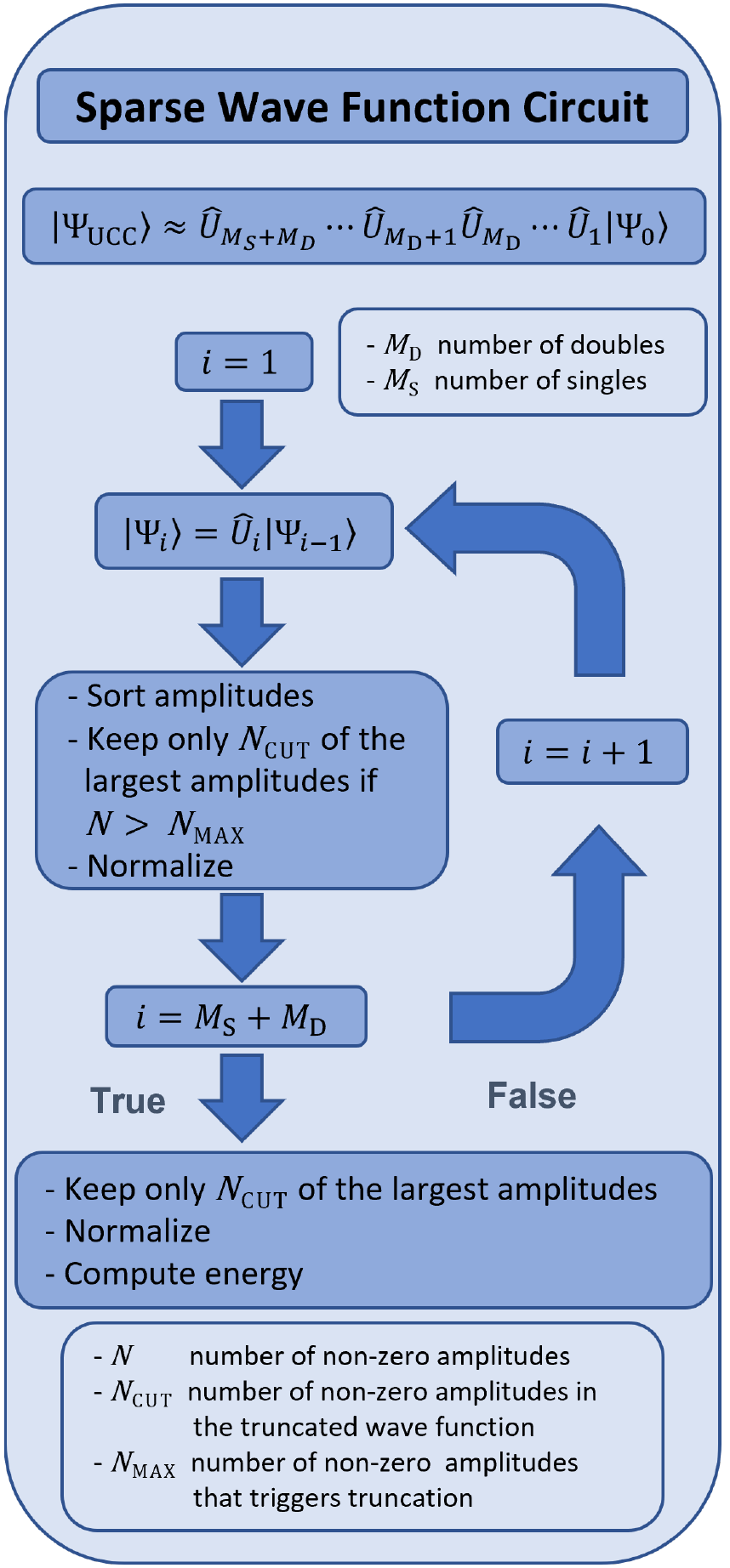}
\caption{The sparse wavefunction circuit for the factorized form of the UCC ansatz where each $\hat{U}_{i}$ is an exponential operator involving a single variational parameter.}
\label{fig:flowchart}
\end{figure}

The unitary coupled cluster (UCC) ansatz $\ket{\Psi_{\mathrm{UCC}}}$ is an exponential form acting on a single reference state $\ket{\Psi_{0}}$: 
\begin{equation}
    \label{eq:ucc}
    \ket{\Psi_{\mathrm{UCC}}} = \exp(\hat{T} - \hat{T}^{\dagger})\ket{\Psi_{0}} 
\end{equation}
where the coupled cluster operator, $\hat{T}$, is defined in terms of the particle-hole excitation and de-excitation operators as
\begin{equation}
    \label{eq:t}
    \hat{T} = \sum_{i}^{\mathrm{occ}} \sum_{a}^{\mathrm{vir}} \theta_{i}^{a}\hat{a}_{i}^{a}
            + \sum_{ij}^{\mathrm{occ}} \sum_{ab}^{\mathrm{vir}} \theta_{ij}^{ab}\hat{a}_{ij}^{ab} + \cdots
\end{equation}
The variational parameters are $\theta_{i}^{a}$ and $\theta_{ij}^{ab}$, and the orbital indices take their conventional meanings: $i$, $j$, $k$, \dots indicate occupied orbitals in the reference state, and $a$, $b$, $c$, \ldots indicate virtual orbitals.
In this work, we truncated $\hat{T}$ to include only $M_{\mathrm{S}}$ single excitation operators and $M_{\mathrm{D}}$ double excitation operators.
We approximate the UCC ansatz with its factorized form given by
\begin{equation}
    \label{eq:ansatz}
  \ket{\Psi_{\mathrm{UCC}}} \approx \hat{U}_{M_{\mathrm{S}}+M_{\mathrm{D}}}\cdots\hat{U}_{M_{\mathrm{D}+1}}\hat{U}_{M_{\mathrm{D}}}\cdots\hat{U}_{1}\ket{\Psi_{0}}  
\end{equation}
where each single UCC factor is given by
\begin{equation}
    \hat{U}_i^a = \exp{\theta_i^a\left(\hat{a}_i^a - \hat{a}_a^i\right)}
\end{equation}
and each double UCC factor is given by
\begin{equation}
    \hat{U}_{ij}^{ab} = \exp{\theta_{ij}^{ab}\left(\hat{a}_{ij}^{ab} - \hat{a}_{ab}^{ij}\right)}
\end{equation}
Reference \cite{Chen:2021:841} provides a simplified form for these individual UCC factors that can be efficiently evaluated on classical hardware. With a few modifications we convert the classical simulation algorithm into a sparse wavefunction circuit optimization for UCC with singles and doubles (UCCSD) that can be performed on classical computers (Figure \ref{fig:flowchart}).

\begin{figure}
\includegraphics[width=3.4in]{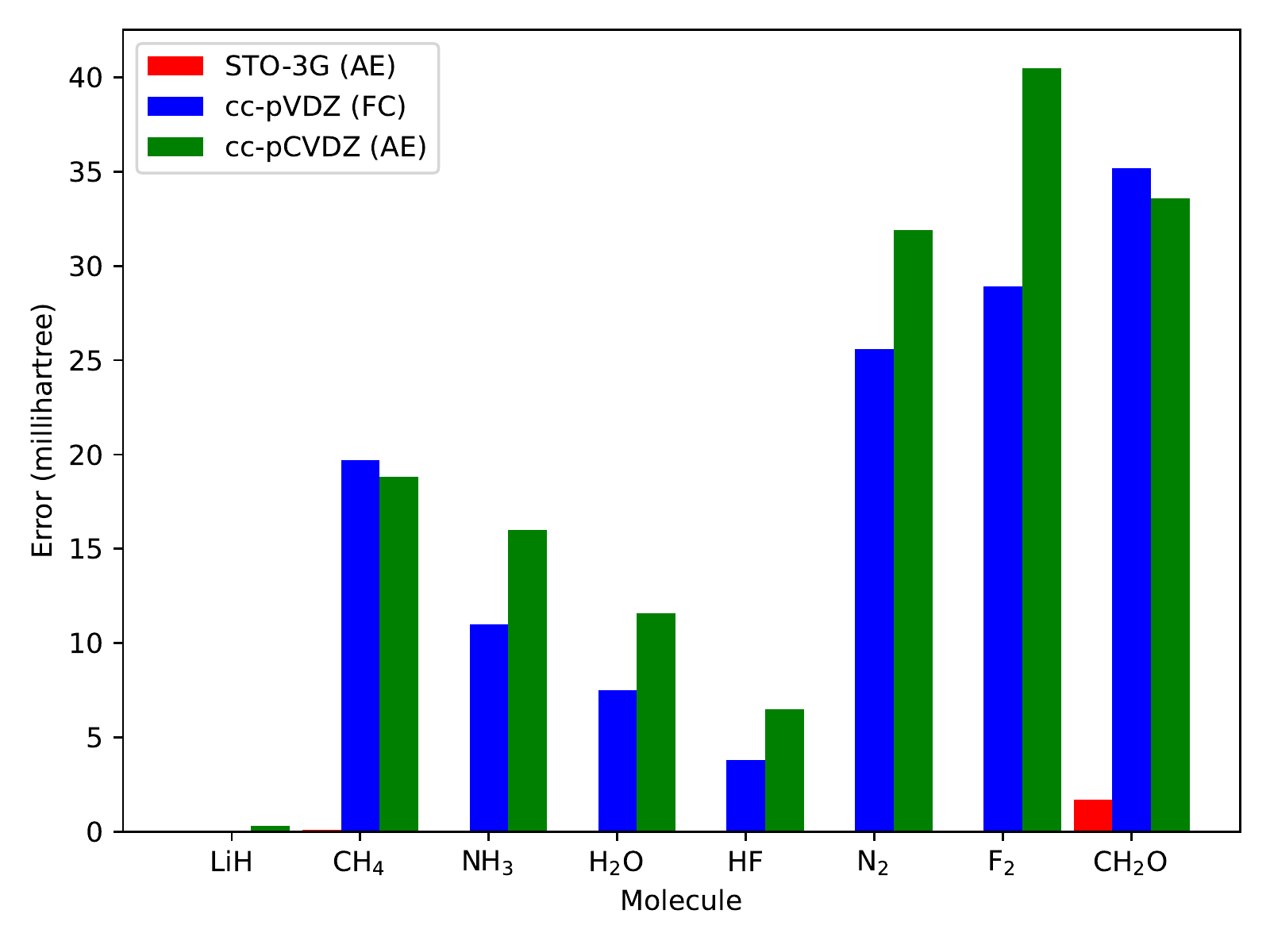}
\caption{Error in the UCCSD energy compared to CCSD(T) using the STO-3G, cc-pVDZ, and cc-pCVDZ basis sets. All-electron (AE) calculations were performed with the STO-3G and cc-pCVDZ basis sets, and the frozen-core approximation (FC) is employed for the cc-pVDZ basis set. The highest energy virtual orbitals are removed in some calculations so that only 64 spin-orbitals are included in the UCCSD and CCSD(T) calculations. For the cc-pVDZ and cc-pCVDZ UCCSD optimizations, $N_{\textrm{CUT}}$ and $N_{\textrm{MAX}}$ are set to 5,000 and 8,000, respectively.}
\label{fig:benchmark}
\end{figure}

We demonstrate our approach by optimizing the UCCSD ansatz for a set of eight molecules using the STO-3G, cc-pVDZ, and cc-pCVDZ basis sets. Figure \ref{fig:benchmark} reports the error in the UCCSD energy compared to the CCSD(T) energy. As a test of the full UCCSD optimization, we did not truncate the ansatz or the wavefunction for the STO-3G calculations. With this minimal basis set, the difference between the UCCSD and CCSD(T) energies is less than 1 millihartree except for CH$_2$O.   The CH$_2$O molecule represents the largest optimization reported for this basis set with 24 qubits, 1,424 variational parameters, and 245,025 Slater determinants in the Hilbert space for this 16 electron system. These results demonstrate that UCCSD provides good agreement with conventional CCSD(T) when there is no truncation of the UCCSD ansatz or wavefunction, at least for a minimal basis set. 

For larger basis sets, the Hilbert spaces are so large as to require truncation of the wavefunction.  For all cases presented here, it is feasible to run the entire UCCSD ansatz but we perform truncation on the ansatz for practical purposes. As an example, CH$_2$O with the cc-pVDZ basis and frozen-core approximation represents a 72 qubit optimization with 34,398 variational parameters and a Hilbert space of $10^{11}$ Slater determinants. To create a standard for our benchmark simulations for the all-electron cc-pCVDZ and frozen-core cc-pVDZ optimizations, we only included 5,000 of the double UCC factors ($M_{\mathrm{D}} = 5,000$) with the largest MP2 amplitudes and all singles as well as set $N_{\textrm{CUT}}$ to 5,000 and $N_{\textrm{MAX}}$ to 8,000. No approximation is needed for the smaller molecules. 
With this approximation, the UCCSD optimizations produce results close to the conventional CCSD(T) correlation energies but, in some cases, underestimate the correlation energy by as much as 40 millihartree. Although we note that these errors are significant, our motivation for this work is to demonstrate that the sparse wavefunction circuit solver can generate reliable parameters for large numbers of qubits (i.e., spin-orbitals) that can be further refined on quantum hardware.  Finding a more accurate and compact wavefunction ansatz is certainly feasible with our approach which we will explore in future work.

\begin{figure}
\includegraphics[width=3.4in]{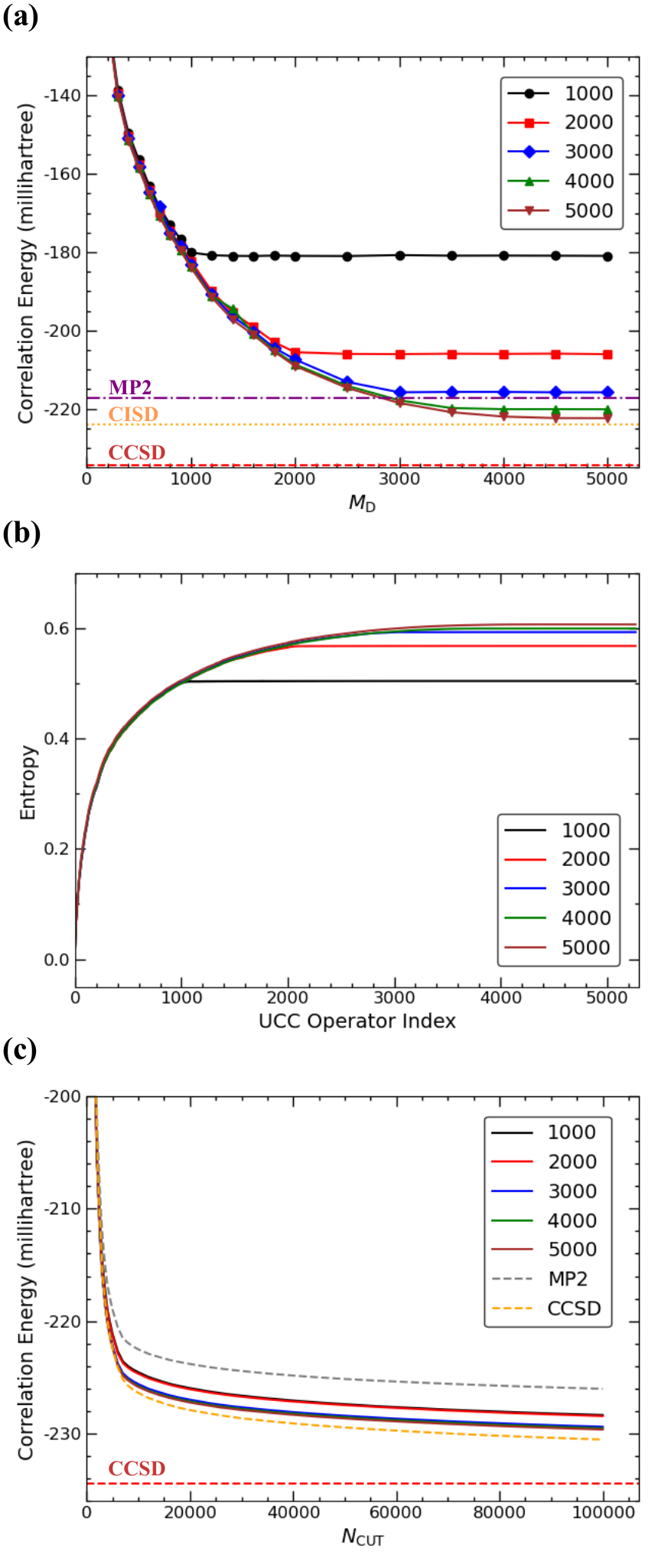}
\caption{(a) Convergence of the correlation energy for NH$_3$ with respect to $M_{\mathrm{D}}$ for $N_{\mathrm{CUT}}$ set to 1,000, 2,000, 3,000, 4,000, and 5,000 and $N_{\mathrm{MAX}}$ set to 8,000. (b) Entropy of the NH$_3$ wavefunction as each UCCSD operator is applied. Parameters were taken from the optimizations with $M_{\mathrm{D}}$ set to 5,000. (c) Correlation energy of NH$_3$ as a function of $N_{\mathrm{CUT}}$ employing optimized UCCSD parameters used in (b). The dashed gray and orange curves employed MP2 and CCSD parameters in the ansatz without optimization.}
\label{fig:nh3}
\end{figure}

In order to understand the performance of UCCSD with both truncation of the ansatz and the wavefunction required for large number of qubits, we closely study the convergence of the correlation energy with respect to the number of double UCC factors ($M_{\mathrm{D}}$) included and $N_{\textrm{CUT}}$ for NH$_3$ with the cc-pCVDZ basis set. Figure \ref{fig:nh3}a illustrates the convergence of the correlation energy with respect to $M_{\mathrm{D}}$ for the NH$_3$ molecule with the cc-pCVDZ basis set. For these optimizations, the highest energy virtual orbital is not included so that there are 32 molecular orbitals corresponding to a 64 qubit optimization using the standard Jordan-Wigner mapping. Our approximate simulator is required since this 10 electron system with 32 molecular orbitals contains $4.1 \times 10^{10}$ determinants in the Hilbert Space.  We  truncate the wavefunction with $N_{\textrm{CUT}}$ set at 1,000, 2,000, 3,000, 4,000, or 5,000 and $N_{\textrm{MAX}}$ set at 8,000. 
An interesting feature of the curves in Figure \ref{fig:nh3}a is that they level-off where $N_{\textrm{CUT}}$ equals $M_{\mathrm{D}}$. This suggests that the optimization becomes over-parameterized when $M_{\mathrm{D}} >N_{\textrm{CUT}}$. Since NH$_{3}$ is a closed-shell molecule, we expect the most important determinants to be the doubly excited determinants, and this is what is seen by analyzing the optimized wavefunction. 
In addition to this leveling-off feature, we do see that the curves are converging as $N_{\textrm{CUT}}$ increases from 1,000 to 5,000 indicating that the correlation energy is converging with respect to the size of the wavefunction and the size of the UCCSD ansatz.

\begin{figure}
\includegraphics[width=3.4in]{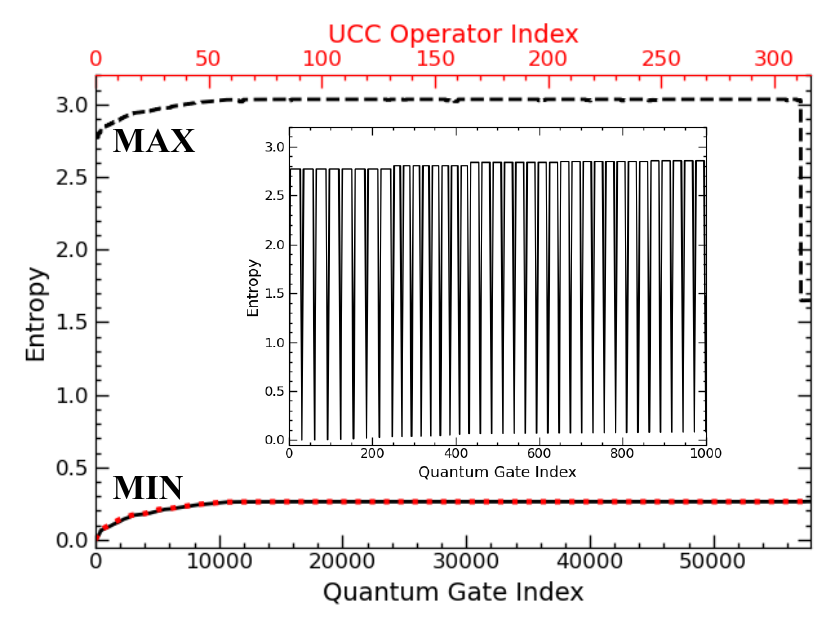}
\caption{Entropy of the wavefunction for NH$_3$ with the STO-3G basis without truncation for the classical UCC ansatz (top axis in red) and the equivalent quantum circuit (bottom axis in black). The solid black line plots the the local minima of the entropy curve, and the dashed black line plots the local maxima of the entropy curve. The dotted red curve (entropy of the classical ansatz) lies on top of the local minima of the quantum circuit demonstrating the equivalence of both methods. Note that only 315 classical operators are equivalent to the over 50,000 quantum gates. The dip in the dashed black curve at the end corresponds to the quantum gates for the single UCC operators. The inset shows the entropy for the first 1,000 quantum gates.}
\label{fig:entropy}
\end{figure}

These results demonstrate the convergence behavior of the UCC energy with respect to the size of the UCC ansatz and the wavefunction and show a definitive classical region of simulability of the UCC ansatz. Next, we explore how the complexity of the wavefunction, which we define in terms of an entropy function, limits the size of the circuit optimization on classical hardware, as opposed to the number of qubits. To measure the complexity of the wavefunction as the UCCSD ansatz is applied, we define the entropy of the wavefunction as 
\begin{equation}
 S = - \sum_{I}^{N_{\mathrm{det}}} \abs{c_{I}}^{2} \ln{\abs{c_{I}}^2}
\end{equation}
where $c_{I}$ is the amplitude for determinant $I$ and $N_{\mathrm{det}}$ is the number of determinants in the wavefunction. Figure \ref{fig:entropy} shows how the entropy changes for NH$_3$ with the STO-3G basis without any truncation of the wavefunction or the ansatz. Here, we compute the entropy using the classical approach presented in this manuscript (red curve) as well as mapping the factorized UCC ansatz to a quantum circuit~\cite{Barkoutsos:2018:022322} (black curves). Here, each UCCSD factor is given an index based on its position in the factorized UCCSD ansatz (Figure {\ref{fig:flowchart}}) with the most important doubles being to the right of the product and the singles to the left. The entropy for the wavefunction with our sparse simulator monotonically increases to a constant value of 0.26 over the 315 classical UCCSD factors. In contrast, the entropy for the quantum circuit oscillates between the values observed from the classical circuit and values above 3.0 over the more than 50,000 quantum gates composing the quantum circuit. This oscillatory behavior is shown in the inset of Figure \ref{fig:entropy} for the initial 1,000 quantum gates. 
In the main part of Figure \ref{fig:entropy}, we only plot the minimum and maximum of each oscillation as two separate curves and we find that the black curve for the minimum entropy of the quantum circuit matches the red entropy curve for the classical circuit. This clearly shows that the wavefunction complexity only covers a small fraction of the Hilbert space and grows as more of the UCC factors are applied.  The quantum circuit has many intermediate circuit elements in which Hilbert space coverage can be large in the computational basis, and our approach circumvents this complexity allowing us to simulate large system sizes.  Various tensor network approaches can also be used to circumvent this type of complexity, but they also incur other computational trade-offs, which is a future research direction.  

Figure \ref{fig:nh3}b shows how the entropy changes as each UCCSD factor with the corresponding optimized parameter is applied for the NH$_3$ using parameters from the optimizations in Figure \ref{fig:nh3}a setting $M_{\mathrm{D}} = 5,000$ and setting $N_{\textrm{CUT}}$ to 1,000, 2,000, 3,000, 4,000, and 5,000. In this plot, we truncate the wavefunction after each factor is applied so that only $N_{\textrm{CUT}}$ determinants are kept. The entropy increases as each UCCSD factor is applied and approximately levels off when the UCC factor index equals $N_{\textrm{CUT}}$. Although the plot appears smooth, there are small variations after each operator due to the truncation of the wavefunction. This shows that the part of the wavefunction that is thrown away makes no significant contribution. This demonstrates that controlling the wavefunction complexity through our truncation mechanism is key for the approximate UCC optimizations described here.


Another important aspect of this study is to examine how effective our approach is for determining starting parameters of simulations on quantum hardware. To this end, we took the optimized parameters used in Figure \ref{fig:nh3}b. With these optimized circuits, we computed the energy without further optimization but allowing the wavefunction to grow beyond $N_{\textrm{CUT}}$ used in the optimization. Figure \ref{fig:nh3}c shows the convergence of the correlation energy for these calculations. This plot shows that even with parameters optimized for smaller wavefunction, we continue to recover correlation energy just by letting the wavefunction grow. This demonstrates that even approximate contributions of the higher excited determinants will recover more correlation energy. Additionally, we performed this calculations with the MP2 amplitudes which were used as starting parameters in the optimization as well as conventional CCSD amplitudes inserted into the UCCSD ansatz. Remarkably, using these non-optimized starting parameters envelops the curves using the optimized UCCSD parameters for all five values of $N_{\textrm{CUT}}$. Since our goal in the near-term is to perform as much of the optimization on classical hardware, this result suggests that it will be advantageous to begin all optimizations with CCSD amplitudes in the weakly correlated limit. Further investigation is needed to validate this claim.




\section{Conclusion} 
\label{sec:conclusions}
In this work, we have described a quantum-inspired sparse wavefunction circuit solver employing the UCCSD ansatz that runs on classical hardware for systems that would require up to 64 qubits on quantum devices, though the restriction to 64 qubits could easily be removed in future work. Although our implementation could be optimized for HPC resources, all calculations reported here could be run on a single computational node. By taking advantage of the sparseness of the wavefunction for the molecular systems studied here, we are able to limit the computational resources needed for these simulations to only the important parts of Hilbert space. These approximations allow the use of double-zeta quality basis sets that are important for describing chemical systems accurately.  
The potential benefits of our approach are numerous and open the path for extensive classical simulations to probe the utility of the UCC ansatz and others for future quantum simulations. 

Our benchmark calculations of the set of eight  molecules shows the performance of the UCCSD ansatz with these approximations compared to conventional coupled cluster theory using double-zeta basis sets. In addition, the study of NH$_3$ demonstrates the convergence behavior of the UCCSD ansatz for recovering correlation energy under these approximations and how controlling the complexity of the wavefunction is key to performing these classical UCCSD optimizations.

\section{Methods}
\label{sec:methods}

Equations 1-5 summarize the UCCSD ansatz employed in this work. Recently, Chen, Cheng, and Freericks~\cite{Chen:2021:841} employed the SU(2) algebra to obtain a simplified form of the individual exponential UCCSD factors in Equation 4 and 5 that can efficiently be evaluated on classical hardware.
This leads to an efficient algorithm inspired by the VQE for UCCSD that can be performed on classical computers. Figure \ref{fig:flowchart} describes how, in combination with truncation of the UCCSD ansatz to include the most important UCC factors based on MP2 amplitudes, the wavefunction complexity can be controlled by discarding insignificant amplitudes after the application of each UCC factor.

We take as the initial values for the variational parameters $\theta$ the MP2 amplitudes for the double excitations, and the single excitations are set to zero. The double UCC factors are ordered based on the magnitude of the initial parameters so that the operator with the parameter with the largest magnitude is applied first, i.e., appears to the right in the factorized form of Equation \ref{eq:ansatz}. To make the calculations manageable, we truncate the number of double UCCSD factors indicated by $M_{\mathrm{D}}$, while all single UCC factors are included in all calculations. In addition to truncating the UCC ansatz, we truncate the wavefunction during application of the UCC ansatz. After application of each UCC factor, if the number of determinants in the wavefunction is larger than $N_{\mathrm{MAX}}$, then we order the determinants based on the magnitude of their amplitudes, and only keep $N_{\mathrm{CUT}}$ of these determinants based on the magnitude of the associated amplitudes. This greatly reduces that complexity of the optimization and allows us to investigate the VQE-inspired approach for large systems. Although we limit the size of the UCC ansatz based on MP2 amplitudes, this approach could be extended to other approaches like ADAPT-VQE \cite{Grimsley:2019:3007}.

The current implementation restricts the number of molecular orbitals to 32, i.~e., 64 spin orbitals. Therefore, for any system considered here with more than 32 molecular orbitals, we only include the 32 lowest-energy molecular orbitals in the correlation energy calculations. Our UCCSD implementation uses PySCF \cite{Sun:2020:024109} to generate the molecular orbitals, molecular integrals, and initial MP2 amplitudes for our UCCSD optimiations. The SLSQP optimizer in the SciPy package \cite{2020SciPy-NMeth} is used in all UCCSD optimizations. Full configuration interaction (FCI), configuration interaction with singles and doubles (CISD), coupled cluster theory with doubles (CCD), coupled cluster theory with singles and doubles (CCSD), and CCSD with perturbative triples [CCSD(T)] energies reported here were computed with PySCF \cite{szabo2012modern,Bartlett:2007:291}. We employ the STO-3G \cite{Hehre:1969:2657}, cc-pVDZ \cite{Dunning:1989:1007}, and cc-pCVDZ \cite{Woon:1995:4572} basis sets. Experimental geometries are used for all molecules and were taken from the Computational Chemistry Comparison and Benchmark DataBase \cite{NIST}.

\section*{Acknowledgments:}
  We are grateful for support from NASA Ames Research Center.
   We acknowledge funding from the NASA ARMD Transformational Tools and Technology (TTT) Project. 
  Calculations were performed as part of the XSEDE computational Project No. TG-MCA93S030 on  Bridges-2 at the Pittsburgh supercomputer center. 

%

\end{document}